# Deep learning framework DNN with conditional WGAN for protein solubility prediction

Xi Han[1], Liheng Zhang[2], Kang Zhou[1,3,*], Xiaonan Wang[1,*]

[1]Department of Chemical and Biomolecular Engineering, National University of Singapore, Singapore, 117585, [2]Department of Computer Science, University of Central Florida, Orlando, FL 32816, USA, [3]Disruptive & Sustainable Technologies for Agricultural Precision, Singapore-MIT Alliance for Research and Technology, Singapore, 138602.

*To whom correspondence should be addressed.

## Abstract

**Motivation:** Protein solubility plays a critical role in improving production yield of recombinant proteins in biocatalyst and pharmaceutical field. To some extent, protein solubility can represent the function and activity of biocatalysts which are mainly composed of recombinant proteins. Highly soluble proteins are more effective in biocatalytic processes and can reduce the cost of biocatalysts. Screening proteins by experiments *in vivo* is time-consuming and expensive. In literature, large amounts of machine learning models have been investigated, whereas parameters of those models are underdetermined with insufficient data of protein solubility. A data augmentation algorithm that can enlarge the dataset of protein solubility and improve the performance of prediction model is highly desired, which can alleviate the common issue of insufficient data in biotechnology applications for developing machine learning models. Here we propose a more accurate prediction model—deep neural networks (DNN) and a data augmentation algorithm—conditional Wasserstein Generative Adversarial Networks (conditional WGAN) for improving prediction of protein solubility.
**Results:** We first implemented a novel approach that a data augmentation algorithm—conditional WGAN was used to improve prediction performance of DNN for protein solubility from protein sequence by generating artificial data. After adding mimic data produced from conditional WGAN, the prediction performance represented by $R^2$ was improved compared with the $R^2$ without data augmentation. After tuning the hyperparameters of two algorithms and organizing the dataset, we achieved a $R^2$ value of 45.04%, which enhanced $R^2$ about 10% compared with the previous study using the same dataset. Data augmentation opens the door to applications of machine learning models on biological data, as machine learning models always fail to be well trained by small datasets.
**Availability:** We present the machine learning workflow as a series of codes hosted on GitHub (https://github.com/xiaomizhou616/DNN_conditional-WGAN). The workflow can be used as a template for the analysis of protein solubility and other datasets.
**Contact:** kang.zhou@nus.edu.sg, chewxia@nus.edu.sg

## 1 Introduction

Developing highly active biocatalysts that can significantly reduce the production cost is a crucial process in pharmaceutical, agricultural and industrial applications. Biocatalysts are mainly composed of various enzymes, which are mostly recombinant proteins. *Escherichia coli* is a bacterium commonly used in genetic engineering to express recombinant proteins (Chan, et al., 2010), however, low activity of those recombinant proteins results in inefficient biocatalytic processes.

Some experimental technologies (i.e. using suitable promoters, optimizing codon usage, or changing culture media, temperature and/or other culture conditions) could enhance the expression of recombinant proteins (Idicula-Thomas and Balaji, 2005; Magnan, et al., 2009), but such empirical optimizations are time-consuming and labor-intensive and more importantly, they often fail due to ambiguous reasons. A more effective solution is highly desired such as an accurate computational model that can predict protein activity from its amino acid sequence directly.

Since activity and solubility of proteins are correlated to some extent, and a dataset of protein solubility is available for exploration, protein solubility has been used as a proxy for protein activity (Han, et al., 2018). The association between amino acid sequence and protein solubility has been investigated by a large amount of machine learning (ML) models and the prediction accuracy is being gradually improved. A simple regression method was first developed by Wilkinson and Harrison (Wilkinson and Harrison, 1991) to predict protein solubility from amino acid sequence, which achieved an accuracy of 0.88 with 81 protein sequences. Subsequently, the accuracy was improved to 0.94 by a logistic regression model with 212 proteins (Diaz, et al., 2010). It was recognized that small dataset is not valid enough for generating a generic model, and then different Support Vector Machine (SVM) models were developed with more than 1000 samples (Niwa, et al., 2009), with 2159 proteins (Agostini, et al., 2012) and with 5692 proteins (Xiaohui, et al., 2014), respectively.

Moreover, various machine learning models were explored, such as decision tree (Christendat, et al., 2000; Goh, et al., 2004; Rumelhart, et al., 1985), Naïve Bayes (Smialowski, et al., 2006), random forest (Fang and Fang, 2013; Hirose, et al., 2011), and gradient boosting machine (Rawi, et al., 2017). Moreover, several software and web servers were also developed for protein solubility prediction, such as ESPRESSO (Hirose and Noguchi, 2013), Pros (Hirose and Noguchi, 2013), PROSOII (Smialowski, et al., 2012), SOLpro (Magnan, et al., 2009) and PROSO (Smialowski, et al., 2006).

Among a large number of databases published in previous studies, eSol database (Niwa, et al., 2009) is the unique one with continuous values of solubility, ranging from 0 to 1. Most machine learning models were developed based on databases divided into two groups, i.e. soluble or insoluble proteins, and only one previous work explored continuous values of protein solubility which are more meaningful to guide experiments in protein engineering (Han, et al., 2018). However, a larger dataset is more desired to train machine learning models fully and improve prediction performance.

Deep learning, a branch of machine learning, has advanced rapidly during the last two decades and demonstrated tremendous progress in bioinformatics (Min, et al., 2017). Applications of deep learning are attracting more and more attention from both academia and industry. Useful insights can be gained by analyzing biological data through various deep learning architectures (e.g., deep neural networks (Hinton, et al., 2012), convolutional neural networks (Krizhevsky, et al., 2012), recursive neural networks (Socher, et al., 2011), recurrent neural networks (Тарасов, 2015) and shallow neural networks (Mikolov, et al., 2013) etc.). Deep neural network (DNN) is well known for its advantages of analyzing data in high dimensions, which is one of the key characteristics of complex bioinformatics data. DNN has shown great potential in exploring patterns or correlations in bioinformatics filed (Aliper, et al., 2016; Xu, et al., 2015; Zhang, et al., 2017) . For predicting protein solubility from protein sequence, a convolutional neural network was investigated recently on a dataset with only binary values of 0 and 1 of protein solubility (Khurana, et al., 2018). However, binary values of protein solubility are not suitable or favored for guiding protein engineering as discussed.

Generative Adversarial Networks (GANs), a state-of-the-art data augmentation algorithm in Artificial Intelligence (AI), have mainly been applied in the fields of image recognition and computer vision. The proposal of GANs can help scientists leverage existing datasets by uncovering patterns in these datasets and reduce hands-on experiments which require large amounts of labor costs. Many variants of GANs have been developed, which enhanced the performance of GANs to some extent. Conditional GAN (CGAN) is the conditional version of GANs, which generates mimic data with class labels (Mirza and Osindero, 2014). Wasserstein GAN (WGAN) uses Wasserstein distance metric rather than Jensen-Shannon distance metric in GANs when the cross-entropy is calculated (Arjovsky, et al., 2017). Conditional WGAN is the conditional version of the Wasserstein GANs (Gulrajani, et al., 2017). Localized GAN (LGAN) uses local coordinate charts to access the local geometry and alleviate mode collapse (Qi, et al., 2018). Recently, various data augmentation algorithms demonstrate a novel and attractive method in generating artificial data in diverse biological areas, such as genes, protein sequences, and drugs. Data augmentation provides a new platform for modeling in biological research limited by small datasets. A novel feedback-loop architecture, called Feedback GAN (FBGAN), was used to generate artificial DNA sequences for optimizing desired antimicrobial properties based on a recurrent neural network classifier (Gupta and Zou, 2018). In addition, a sequence generating framework, called SeqGAN, was proposed for generating sequences of discrete tokens to reduce the cost of extensive experiments on synthetic data (Yu, et al., 2017). Moreover, some architectures of conditional WGAN shows great promise in datasets with labels, which generate mimic data similar to real data, and explore the relationship between features and labels at the same time. For example, image synthesis has been studied in image quality assessment (Odena, et al., 2016) and image-to-image translation problems were also solved by conditional adversarial networks effectively (Isola, et al., 2017; Zhu, et al., 2017). However, conditional WGAN has not been investigated to solve the problem of protein solubility, which also has some feature protein sequence and a target response protein solubility. Data augmentation which can alleviate the problem of lacking sufficient data is always highly desired in biological fields.

In the present study, we built a conditional WGAN to generate artificial data to improve the performance of a deep learning framework DNN that predicts protein solubility in continuous values from 0 to 1 with the eSol dataset. DNN was explored as the prediction model for protein solubility for the first time, which is more suitable for our relatively small dataset, compared with the convolutional neural network used before (Khurana, et al., 2018). For all experiments in our study, conditional GAN successfully enhanced the regression evaluation metric $R^2$ based on DNN model by data augmentation, which shows a novel and promising approach for enlarging biological datasets without time-consuming experiments *in vivo*. We compared our approach with previous work predicting continuous values of solubility from protein sequence (Han, et al., 2018), and demonstrated that the proposed method has improved $R^2$ from 41.07% to 45.04%.

## 2 Methods

### 2.1 Protein database

All the protein solubility information used in the study was obtained from eSol database (Niwa, et al., 2009). For collecting protein solubility in this dataset, chaperone-free protein expression technology was used at first to produce proteins *in vitro* and complete *E. coli* ORF library (ASKA library) consists all predicted Open Reading Frame (ORF) (Kitagawa, et al., 2005). Synthesized proteins were then separated into soluble and insoluble fractions by using centrifugation after translation, and the proteins in both fractions were quantified using SDS-PAGE. Finally, solubility is recorded as the percentage of the supernatant protein quantity among the total protein quantity. All the samples from the database were used except the samples containing no sequence information, or poorly determined sequence (containing N instead of A, T, C, or G, or multiple stop-codons). 25 samples were excluded from 3173 samples of the eSol database, and the remaining 3148 proteins with protein sequence and protein solubility were the whole dataset in our study.

### 2.2 Features

Different features were extracted from sequences of proteins by using protr package (Xiao, et al., 2014) within R software, which generates various numerical representation schemes. Comparing the prediction performance of different descriptors, Amino Acid Composition Descriptor in protr package achieved the best performance according to the previous study (Han, et al., 2018), which was used in our work. For this descriptor, the amino acid composition of 20 amino acids was listed in first 20 columns in our dataset.

### 2.3 Training flowsheet

Fig. 1 shows the whole workflow of our work. Data pre-processing was conducted at first to transform characters of protein sequence into numerical values of the amino acid composition by Amino Acid Composition Descriptor. In addition, values of protein solubility were converted into continuous values within range 0–1. After extracting a well-organized dataset including the amino acid composition *x* in first 20 columns and protein solubility *y* in the last column from raw data, it was taken as inputs to train DNN model.

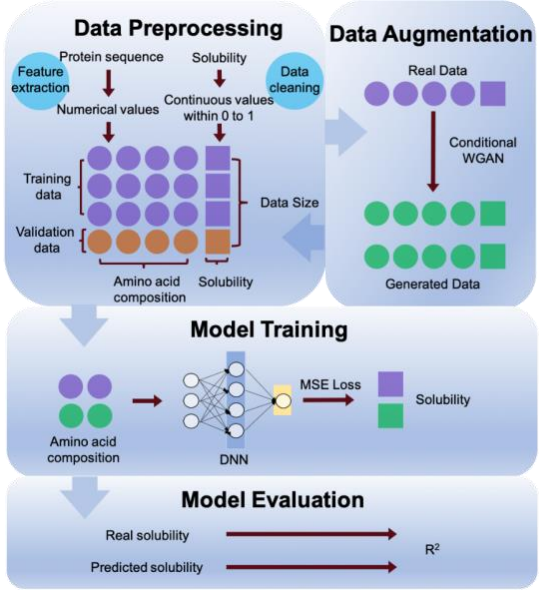

**Fig. 1. The workflow of our study**

The original dataset including 3148 proteins was divided into a 4-fold dataset with the same size according to the original order in the dataset. Each group was twenty-five percent of the original dataset (787 proteins) and was selected as validation data once to test the performance of DNN trained by remaining training data (2361 proteins) in the original dataset. In Table 3, all the 4 groups were tested 5 times to reduce the variance and average results were used as evaluation metrics. In the process of training DNN, validation data were used to test the model performance by calculating $R^2$. Subsequently, the training data were taken as inputs of conditional WGAN to produce doubled dataset including the training data in the original dataset and generated data in the same size to train DNN again. A new $R^2$ value was measured on the validation data which were held out in the process and compared with the original $R^2$. The difference between the two $R^2$ reflects the performance of the data augmentation algorithm conditional WGAN.

### 2.4 Machine learning model—DNN

A DNN is a neural network with a certain level of complexity, i.e., a neural network with multiply hidden layers (Bengio, 2009; LeCun, et al., 2015). The supervised machine learning algorithm DNN was applied to our dataset to predict protein solubility from the amino acid sequence. For continuous values of solubility, the regression algorithm of DNN was used in our study.

### 2.5 Data augmentation algorithm—conditional WGAN

Generative Adversarial Networks are a class of ML algorithms used for data augmentation. GANs consist of two neural networks, Generator and Discriminator, which are trained in opposition to each other. The generator *G* takes random noise data *z* from probability distribution $P_z(z)$ as inputs and synthesized data *G(z)* as outputs. The discriminator *D* receives data generated from *G* and real data from training dataset as inputs and produces a probability distribution to distinguish whether the data are artificial or real as outputs. In Equation (1), the discriminator is trained to maximize the log-likelihood it assigns to the correct data source (termed as max(*V*)), whereas the generator is trained to minimize max(*V*). This adversarial game between *G* and *D* will finish until *D* cannot distinguish the generated data and real data anymore.

$$\min_G \max_D V(D,G) = E_{x \sim P_{data}(x)}[\log D(x)] + E_{z \sim P_{z(z)}}[\log(1 - D(G(z)))] \quad (1)$$

Our goal is to generate data with expected protein solubility to augment real data. We use conditional WGAN (Fig. 2), a variant of the GANs, to achieve this. The input of generator consists of a noise vector *z* with 20 dimensions and a latent label $\tilde{y}$. The latent label serves as conditions which were expected to be the protein solubility of the generated data, and follows the uniform distribution $P_f$ from 0 to 1. The generator is trained to produce data $\tilde{x}$ conditioned on the expected protein solubility $\tilde{y}$ to "fool" the discriminator, $G(z, \tilde{y}) \rightarrow \tilde{x} \sim P_g$. In addition to distinguish real and generated data, we introduce an auxiliary regressor (Odena, A., et al. 2016) to predict the protein solubility, $D: x \rightarrow \{D_{adv}, D_{reg}\}$.

An adversarial loss is adopted to make the generated data more realistic. To stabilize the training, we use the Wasserstein GAN objective with gradient penalty (Arjovsky, M., et al. 2017, Gulrajani, I., et al. 2017) formulated as

$$L_{adv} = E_{x \sim P_r}[D_{adv}(x)] - E_{\tilde{x} \sim P_g}[D_{adv}(\tilde{x})] - \lambda E_{\hat{x} \sim P_{\hat{x}}}\left[\left(||\nabla_{\hat{x}} D_{critic}(\hat{x})||_2 - 1\right)^2\right], \quad (2)$$

where $D_{adv}$ denotes the WGAN value, $\hat{x}$ is the data sampled uniformly along a straight line between a pair of real data *x* and generated data $\tilde{x}$, $\lambda$ is the hyperparameter for gradient penalty and equals 1 in our experiments. The adversarial loss is maximized while training the discriminator *D*, but minimized while training the generator *G*. To generate data with conditioned protein solubility, we introduce an auxiliary regressor on *D*. Given a real data *x*, the discriminator outputs $D_{reg}$ which regresses the corresponding protein solubility *y* of the real data. It learns the correlation between the protein sequence and the protein solubility. The regression loss for real data is defined as

$$L_{reg}^r = E_{x \sim P_r, y \sim P_y} ||D_{reg}(x) - y||_2^2. \quad (3)$$

During the training of generator, the auxiliary regressor provided gradients to constrain the protein solubility of the generated data to be the expected values, which was provided as the condition in the inputs of the generator. Thus, the regression loss for generated data is formulated as

$$L_{reg}^f = E_{\tilde{x} \sim P_g, \tilde{y} \sim P_f} ||D_{reg}(\tilde{x}) - \tilde{y}||_2^2. \quad (4)$$

The total objectives to train *G* and *D* are

$$L_G = L_{adv} + \alpha L_{reg}^f, \quad (5)$$

$$L_D = -L_{adv} + \beta L_{reg}^r, \quad (6)$$

respectively, where $\alpha$ and $\beta$ balance the importance between the adversarial loss and the regression loss. We use $\alpha = 1$ and $\beta = 1$ in all the experiments. After training the generator and the discriminator iteratively for many epochs, the discriminator cannot distinguish generated data from the generator and real data from original dataset anymore and the generated data were the final outputs from conditional WGAN.

After the convergence of the conditional WGAN, we augment the real data with the generated data to train a DNN regressor *R* with loss function

$$\min_R E_{x \sim P_r \cup P_g, y \sim P_y \cup P_f} ||R(x) - y||_2^2. \quad (7)$$

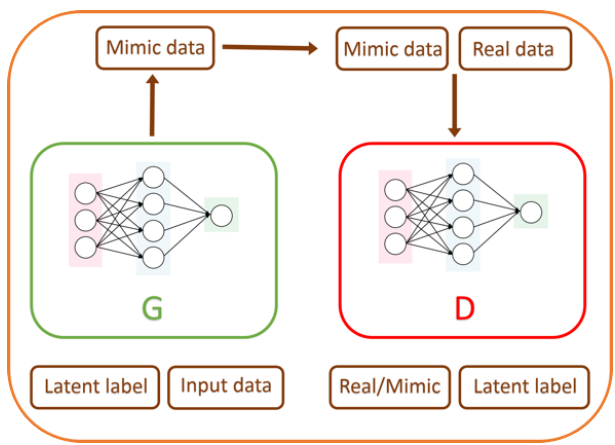

**Fig. 2.** Diagram of conditional WGAN

The architectures of DNN and conditional WGAN are illustrated in Table 1 and 2 respectively. Especially in the discriminator, layer 4(1) and layer 4(2) are two branches in a parallel fashion, producing $D_{adv}$ and $D_{reg}$ respectively.

### 2.7 Evaluation metrics

According to the previous study (Han, et al., 2018), continuous values of predicted protein solubility from regression models are better for guiding protein engineering since the improvement of protein solubility after mutation can be observed easily and several samples for experimental validation can be selected according to the exact values of protein solubility. Therefore, several commonly used evaluation metrics of the regression model, such as the coefficient of determination ($R^2$) were used to quantitatively measure the performance of DNN and conditional WGAN. The $R^2$ of DNN represents the performance of prediction and the improvement of $R^2$ after using conditional WGAN reflects the performance of data augmentation.

## 3 Results

### 3.1 Tuning hyperparameters of DNN and conditional WGAN

After narrowing down search space firstly, grid search was used to tune the hyperparameters in neural networks including the number of layers, neurons and activation function. The architectures we used for DNN and conditional WGAN were listed in Table 1 and 2. For reducing the variance of DNN and conditional WGAN, both models were trained 5 times for the same hyperparameters and the average $R^2$ was taken as the evaluation metrics. It can be seen from Table 3 that conditional WGAN improved the performance of DNN by doubling the training data in the original dataset through data augmentation. The default size of generated data is the same as the training data. The $R^2$ of fold 1, 2, 3, and 4 means the performance for DNN and conditional WGAN when fold 1, 2, 3, and 4 was taken as the validation data and the remaining data was the training data respectively. The results of our 4-fold cross validation were plotted in Fig. 3 and the error bars show the variance of models trained for five times. It was observed that $R^2$ was improved 1.5%, 2.4%, 1.8%, and 4.0% in average after data augmentation when the fold 1, 2, 3, and 4 was selected as validation data, respectively. The difference of improvement in $R^2$ was mainly due to the uneven data distribution in the original dataset, since the folds were divided according to the order of original dataset. After tuning the hyperparameters, data argumentation has successfully enhanced the performance of prediction of protein solubility by enlarging the dataset, which may achieve it by learning more accurate data distribution in the original data.

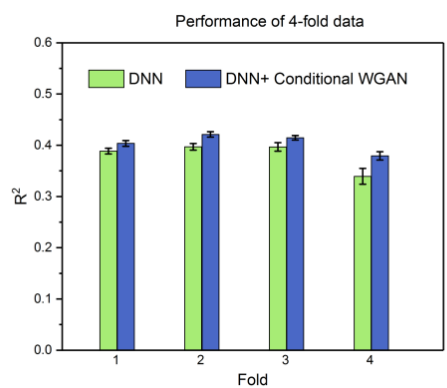

**Fig. 3.** Performance of DNN with and without data augmentation for 4-fold data

**Table 1**. Architecture of DNN

| Linear Layer | 1 | 2 | 3 | 4 |
|---|---|---|---|---|
| Neuron | (20,512) | (512,512) | (512,512) | (512,1) |
| Activation function | ReLU | ReLU | ReLU | Sigmoid |
| Loss function | Equation (7) | | | |

**Table 2**. Architecture of conditional WGAN

| | Linear Layer | 1 | 2 | 3 | 4 | – |
|---|---|---|---|---|---|---|
| Generator | Neuron | (21,512) | (512,512) | (512,512) | (512,20) | |
| | Activation function | ReLU | ReLU | ReLU | Sigmoid | |
| | Loss function | Equation (5) | | | | |
| | Linear Layer | 1 | 2 | 3 | 4(1) [a] | 4(2) [a] |
| Discriminator | Neuron | (20,512) | (512,512) | (512,512) | (512,1) | (512,1) |
| | Activation function | ReLU | ReLU | ReLU | – | Sigmoid |
| | Loss function | Equation (6) | | | | |

[a] 4(1) and 4(2) are two branches on the top of layer 3 in a parallel fashion. 4(1) together with layer 1–3 performs as the critic to distinguish real data and generated data. 4(2) is introduced as an auxiliary regressor. It learns the relation between the real sequence data and its corresponding protein solubility.

**Table 3.** $R^2$ of DNN with and without conditional WGAN

| Dataset | | Algorithm | 1 | 2 | 3 | 4 | 5 | Average |
|---|---|---|---|---|---|---|---|---|
| Fold 1 | | DNN | 0.3952 | 0.3888 | 0.3890 | 0.3795 | 0.3906 | 0.3886 |
| | | DNN+ conditional WGAN | 0.4029 | 0.4037 | 0.4128 | 0.3986 | 0.4001 | 0.4036 |
| Fold 2 | | DNN | 0.3969 | 0.3919 | 0.4062 | 0.3898 | 0.4002 | 0.3970 |
| | | DNN+ conditional WGAN | 0.4193 | 0.4162 | 0.4224 | 0.4297 | 0.4182 | 0.4212 |
| Fold 3 | | DNN | 0.3983 | 0.4021 | 0.4018 | 0.3991 | 0.3822 | 0.3967 |
| | | DNN+ conditional WGAN | 0.4095 | 0.4165 | 0.4205 | 0.4107 | 0.4154 | 0.4145 |
| Fold 4 | | DNN | 0.3312 | 0.3521 | 0.3233 | 0.3592 | 0.3312 | 0.3394 |
| | | DNN+ conditional WGAN | 0.3777 | 0.3783 | 0.3935 | 0.3728 | 0.3747 | 0.3794 |
| Balanced dataset | 100% data | DNN | 0.4205 | 0.4233 | 0.4285 | 0.4292 | 0.4277 | 0.4258 |
| | | DNN+ conditional WGAN | 0.4394 | 0.4369 | 0.4435 | 0.4353 | 0.4365 | 0.4383 |
| | 50% data | DNN+ conditional WGAN | 0.4354 | 0.4274 | 0.4462 | 0.4343 | 0.4348 | 0.4356 |
| | 200% data | DNN+ conditional WGAN | 0.4381 | 0.4401 | 0.4379 | 0.4286 | 0.4335 | 0.4356 |
| | 0.3–0.7 [a] | DNN+ conditional WGAN | 0.4138 | 0.4100 | 0.4283 | 0.4135 | 0.4332 | 0.4198 |
| | 0.3–0.7 [b] | DNN+ conditional WGAN | 0.4255 | 0.4204 | 0.4295 | 0.4315 | 0.4168 | 0.4247 |
| | Sigmoid | DNN | 0.4392 | 0.4436 | 0.4423 | 0.4393 | 0.4297 | 0.4388 |
| | **Sigmoid** | **DNN+ conditional WGAN** | **0.4477** | **0.4500** | **0.4531** | **0.4496** | **0.4515** | **0.4504** |

[a] The conditions of protein solubility were 0–1 in the training process, but the generated data in the augmentation process were limited in the range 0.3–0.7.

[b] The conditions of protein solubility were limited in the range 0.3–0.7 in both the data training and augmentation processes.

### 3.2 Improving model performance by pre-processing the dataset

To further improve the prediction performance of DNN and data augmentation performance of conditional WGAN, the characteristics of our dataset were investigated. This part was critical as data pre-processing is a significant part in data mining and well-organized datasets can improve the quality of modeling. Three major steps, including balancing our dataset by selecting equal data size in different ranges of protein solubility, balancing our dataset by adding data generated from conditional WGAN in the specific range of protein solubility, and enlarging dataset by adding different data size of generated data were explored one by one in our study.

In the training process of DNN, it was observed that the loss of training data was low, whereas the loss of validation data was high. Overfitting occurred and the low loss of training data means the features can be learned by the proposed architecture of DNN. In addition, the performances of different groups were different, which means the original dataset was not distributed evenly and a more balanced dataset was needed to eliminate overfitting and the unbalanced issue. Therefore, we explored the data distribution of protein solubility in our original dataset and plotted it in Fig. 4. It can be found that the values of solubility were not evenly distributed in different ranges between 0–1 and there were fewer points in the range 0.3–0.7 of solubility. The original dataset was divided into 10 sub-groups for solubility within the range of 0–1, using 0.1 as the step size. Then 75% data of each group were taken out randomly and combined as training data, and the remaining data were validation data. It can be seen from Table 3, the balanced dataset achieved better results with a $R^2$ value of 0.4258. The $R^2$ was further improved by 1.25% after data augmentation with conditional WGAN.

In most classification tasks for this dataset, points of solubility in the range of 0.3–0.7 were discarded (Fang and Fang, 2013; Samak, et al., 2012) and fewer points may lead to undesired prediction performance of machine learning models. Therefore, the balanced dataset was enhanced further by generating more points within range 0.3–0.7 by conditional WGAN. Firstly, the ratio of proteins with solubility within 0.3–0.7 to proteins with solubility within 0–1 was calculated. If it was distributed evenly in the range of 0–1, the ratio of protein within 0.3–0.7 should be 40% rather than 30%. Therefore, 420 points with solubility within 0.3–0.7 were generated from conditional WGAN to balance our dataset further. In Table 3, the conditions of protein solubility were limited to the range of 0.3–0.7 ("0.3–0.7") where there were fewer samples in the original dataset. However, this method did not improve the prediction performance. Both taking training data in all ranges of solubility evenly and generating mimic data in specific regions of original dataset are to balance our dataset, whereas the former helped the model learn various samples fully and the latter used the model to produce data in the range of 0.3–0.7 which were learned less in the training process. For the second method, insufficient training for samples with solubility within 0.3–0.7 may result in the disability of the conditional WGAN to generate realistic data, especially only generating mimic data in this region. This may cause that the former improves the model effectively and the latter is not helpful.

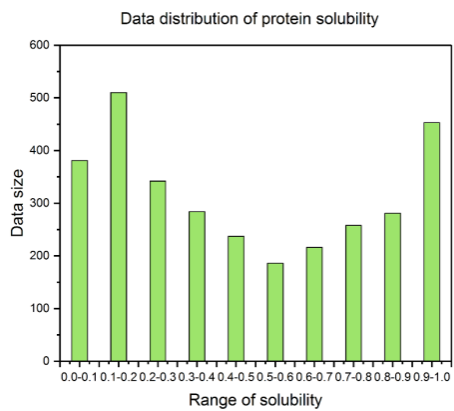

**Fig. 4.** Data distribution of protein solubility in the original dataset

For the previous exploration, the generated data from conditional WGAN had the same size as the training data ("100% data"). Other data size was also tried here such as adding half of generated data ("50% data") and double data size ("200% data") of the training data. It can be observed in Table 3 that there was no obvious improvement compared with the $R^2$ value of 0.4383. The data size of generated data did not influence the data augmentation substantially.

### 3.3 Optimizing activation functions of conditional WGAN

A sigmoid activation function enables the output of regression for protein solubility from protein sequence within the range of 0 to 1, but the boundary values 0 and 1 excluded. Previously protein solubility in our dataset included points with solubility of 0 or 1, therefore the sigmoid activation was not adopted in the model. However, a sigmoid function is commonly utilized even though there exists 0 and 1 in the original dataset. Therefore, a sigmoid function was added here and the $R^2$ was improved to 0.4504 according to Table 3. In addition, the convergence of regression became faster with the sigmoid function. The solubility could infinitely approach 0 or 1, although the exact boundary values were not included. Compared with our previous work, the $R^2$ was improved from 0.4107 to 0.4504 (a close to 10% improvement) after combining the deep learning prediction model DNN and the data augmentation algorithm conditional WGAN.

## 4 Discussion

We have demonstrated the value of our work in two parts: 1) proposing a promising and novel approach to enlarge datasets in biological fields, and 2) improving the model prediction performance for protein solubility. For the first part, generating and collecting a large amount of data by *in vivo* experiments is time-consuming and expensive, which causes a common problem in biotechnology field that the dataset is always too small compared with other fields that use data mining. The machine learning models cannot be fully trained by inadequate volume of data. Data augmentation for data in the biological field is highly desired and also very challenging. First, application of data augmentation in biological datasets is more difficult for evaluation compared with image recognition where data visualization is easier. The characteristics of biological data result in a difficult model evaluation process and few applications of data augmentation algorithm on biological data. Second, using data augmentation for protein sequence is more complex than DNA sequence with only four characters (Gupta and Zou, 2018). Third, it is more complicated to implement conditional WGAN which needs to simulate not only the real data distribution but also the relationship between features and conditions compared with WGAN (Gupta and Zou, 2018). In this work, we have successfully developed a conditional WGAN model to improve the deep learning framework DNN for predicting protein solubility from protein sequence by generating artificial data. For the second part, compared with the previous work (Han, et al., 2018), a more accurate modeling method was achieved by DNN and conditional WGAN and $R^2$ was enhanced about 10%, which may influence the success rate of experimental results in protein engineering. Experimental applications would be based on the prediction results of our model. Therefore, a more accurate model is more effective to develop biocatalysts with high activity that can reduce the production cost significantly in biocatalytic processes. In future work, we would also like to further explore DNN and conditional WGAN by inputting protein sequence as features directly rather than amino acid composition. The largest length of the protein sequences in our dataset is about 200 and one-hot encoding can be used to deal with the sequence data in the inputs. In this way, more information can be included in the model training not only amino acid composition, such as protein sequence, which is very meaningful since mutation for improving protein solubility can be more accurate according to the sequence. In addition, the architecture of DNN and conditional WGAN can be tuned further to improve prediction performance. Linear layers in DNN and conditional WGAN can be replaced by other architectures such as convolutional layers. Finally, DNN can be combined with conditional WGAN into a whole model to tune hyperparameters and the difference between $R^2$ without and with data augmentation can be used as the loss function of the whole synthesized model to further enhance data augmentation.

## Funding

This work was supported by MOE Research Scholarship, MOE Tier-1 grant (R-279-000-452-133) and NRF CREATE Program "Disruptive & Sustainable Technologies for Agricultural Precision" in Singapore.

*Conflict of Interest:* none declared.